\documentclass[10pt]{article}

\usepackage{latexsym,amsmath,amscd,amssymb,graphics}

\usepackage{enumerate}

\usepackage{graphicx}

\usepackage[colorlinks]{hyperref}

\usepackage{url}

\usepackage[all]{xy}

\makeatletter

\@addtoreset{figure}{section}

\def\thefigure{\thesection.\@arabic\c@figure}

\def\fps@figure{h, t}

\@addtoreset{table}{bsection}

\def\thetable{\thesection.\@arabic\c@table}

\def\fps@table{h, t}

\@addtoreset{equation}{section}

\makeatother

\begin{document}

\title{An Example of Banach and Hilbert manifold~:\\ the universal Teichm\"uller space}
\author{A.~B.~Tumpach}

\maketitle

\newtheorem{theorem}{Theorem}[section]

\newtheorem{definition}[theorem]{Definition}

\newtheorem{lemma}[theorem]{Lemma}

\newtheorem{remark}[theorem]{Remark}

\newtheorem{proposition}[theorem]{Proposition}

\newtheorem{corollary}[theorem]{Corollary}

\newtheorem{example}[theorem]{Example}

\def\below#1#2{\mathrel{\mathop{#1}\limits_{#2}}}





\section{Motivations}

\paragraph{$H^s$-Diffeomorphisms groups of the circle.}

For $s>3/2$, the group $\operatorname{Diff}^s(S^1)$ of Sobolev class $H^s$ 
diffeomorphisms 
of the circle is a 
$\mathcal{C}^\infty$-manifold modeled on the space of $H^s$-section of the
tangent bundle $TS^1$ (\cite{Ebin}), or equivalently on the space of real
$H^s$-function on $S^1$. It is a topological group in the sense that the
mutiplication $(f, g)\mapsto f\circ g$ is well-defined and continuous, 
the inverse $f \mapsto f^{-1}$ is continuous, the left translation $L_\gamma$
by $\gamma\in\operatorname{Diff}^s(S^1)$ applying $f$ to $\eta\circ f$ is
continuous, and the right translation 
$R_\gamma$ 
by $\gamma\in\operatorname{Diff}^s(S^1)$ applying $f$ to $f\circ \eta$ is
smooth. These results are consequences of the Sobolev Lemma which states that
for a compact manifold of dimension $n$, 
the space of $H^s$-sections of a vector bundle $E$ over $M$ is contained, for
$s>k +n/2$, in the
space of $\mathcal{C}^k$-sections, and that 
the injection $H^s(E)\hookrightarrow
\mathcal{C}^k(E)$ is continuous. In particular, for  $s>3/2$,
$\operatorname{Diff}^s(S^1)$ is the intersection of the space of 
$\mathcal{C}^1$-diffeomorphisms of the circle with the space $H^s(S^1, S^1)$ 
of $H^s$ maps
from $S^1$ into itself. Hence $\operatorname{Diff}^s(S^1)$ is an 
open set of $H^s(S^1, S^1)$.

For the same reasons, the subgroup of 
$\operatorname{Diff}^s(S^1)$ preserving
three points in $S^1$, say $-1, -i$ and $1$, is, for $s>3/2$, 
a $\mathcal{C}^\infty$ manifold and a topological group
modeled on the space of $H^s$-vector fields which vanish on $-1, -i$ and $1$.

One may ask what happens for the critical value $s = 3/2$ and look for a
group with some regularity and a manifold structure such that 
the tangent space at the
identity is isomorphic to the space of $H^{\frac{3}{2}}$-vector fields
vanishing at $-1, -i$ and $1$ (or equivalently on 
any codimension 3 subspace of $H^{\frac{3}{2}}$). 
The universal Teichm\"uller space $T_0(1)$
defined below will verify these conditions.
 

\paragraph{$\operatorname{Diff}^+(S^1)$ as a group of symplectomorphisms.}

Consider the Hilbert space 
$\mathcal{V} = H^{\frac{1}{2}}(S^1, \mathbb R)/\mathbb R$
of real valued $H^{\frac{1}{2}}$ functions with mean-value zero. 
 Each element $u\in \mathcal{V}$ can be written as

\[
u(x)=\sum_{n\in\mathbb{Z}}u_ne^{inx}\quad\text{with}\quad u_0=0,\; u_{-n}=\overline{u_n}\quad\text{and}\quad \sum_{n\in\mathbb{Z}}|n||u_n|^2<\infty.
\]
Endow $\mathcal{V}$ with the symplectic form
\[
\Omega(u,v)=\frac{1}{2\pi}\int_{S^1}u(x)\partial_xv(x)dx
=-i\sum_{n\in\mathbb{Z}}nu_n\overline{v_n},
\]
The group of orientation preserving
$\mathcal{C}^\infty$-diffeomorphisms of the circle acts on $\mathcal V$ by 
\[
\varphi\cdot f = f \circ \varphi - \frac{1}{2\pi}\int_{S^1} f\circ \varphi, 
\]
preserving the symplectic form $\Omega$.
Note that the previous action is well-defined for any orientation preserving
homeomorphism of $S^1$. Therefore one may ask what is the biggest subgroup of
the orientation preserving homeomorphisms of the circle which preserves
$\mathcal V$ and $\Omega$. The answer is the group of quasisymmetric
homeomorphisms of the circle defined below 
(Theorem~3.1 and Proposition~4.1 in \cite{NaSu1995}).


\paragraph{Teichm\"uller spaces of compact Riemann surfaces.}

Consider a compact Riemann surface $\Sigma$. The Teichm\"uller space 
$\mathcal{T}(\Sigma)$ of
$\Sigma$ is defined as the space of complex structures on $\Sigma$ modulo the
action by pull-back of the
group of diffeomorphisms which are homotopic to the identity. It can be
endowed with a Riemannian metric, called the Weil-Petersson metric, which is
not complete. A point beyond which a geodesic can not be continued corresponds
to the collapsing of a handle of the Riemann surface (\cite{Tromba}), 
hence
yields to a Riemann surface with lower genus. One can ask for a Riemannian
manifold in which all the Teichm\"uller spaces of compact Riemann surfaces
with arbitrary genus inject isometrically. The answer will be the universal
Teichm\"uller space endowed with a Hilbert manifold structure and 
its Weil-Petersson metric (\cite{TaTe2004}). 

  
\section{The universal Teichm\"uller space}


\paragraph{Quasiconformal and quasisymmetric mappings.} Let us give some
definitions and basic facts on quasiconformal and quasisymmetric mappings.
\begin{definition}{\rm
An orientation preserving homeomorphim $f$ of an open subset $A$ in $\mathbb
C$ is called quasiconformal if the following conditions are satisfied.
\begin{itemize}
\item $f$ admits distributional derivatives $\partial_z f$, $\partial_{\bar
    z}f\in L^1_{loc}(A,\mathbb C)$~;
\item there exists $0\leq k < 1$ such that $|\partial_{\bar z}f(z)|\leq k
  |\partial_zf(z)|$ for every $z \in A$.
\end{itemize}
Such an homeomorphism is said to be $K$-conformal, where 
$K = \frac{1 + k}{1 - k}.$}
\end{definition}
For example, $f(z) = \alpha z + \beta\bar z$ with $|\beta|<|\alpha|$
is $\frac{|\alpha|+|\beta|}{\alpha-|\beta|}$-quasiconformal.

\begin{theorem}[\cite{Le1987}]
An orientation preserving
homeomorphism $f$ defined on an open set $A\subset\mathbb C$ is
quasiconformal if and only if it admits distributional derivatives 
$\partial_z f$, $\partial_{\bar
    z}f\in L^1_{loc}(A,\mathbb C)$ which satisfy
$$
\partial_{\bar z} f(z) = \mu(z) \partial_z f(z), \quad z\in A 
$$
for some $\mu\in L^{\infty}(A, \mathbb C)$ with $\|\mu\|_\infty < 1$.
\end{theorem}
The fonction $\mu$ appearing in the previous theorem is called the Beltrami
coefficient or the complex dilatation of $f$. Let $\mathbb D$ denote the open 
unit disc in $\mathbb C$.

\begin{theorem}[Ahlfors-Bers]
Given $\mu\in L^{\infty}(\mathbb D, \mathbb C)$ with $\|\mu\|_\infty<1$, there exists
a unique quasiconformal mapping $\omega_{\mu}~:\mathbb D\rightarrow\mathbb D$
with Beltrami coefficient $\mu$, extending continuously to 
$\overline{\mathbb D}$, and fixing $1, -1, i$. 
\end{theorem}

\begin{definition}{\rm
An orientation preserving homeomorphism $\eta$ of the circle $S^1$ is called
quasisymmetric if there is a constant $M> 0$ such that for every $x\in\mathbb
R$ and every $|t|\leq \frac{\pi}{2}$
$$
\frac{1}{M}\leq \frac{\tilde\eta(x+t)-\tilde\eta(x)}{\tilde\eta(x)-\tilde\eta(x-t)}\leq M,
$$
where $\tilde\eta$ is the increasing homeomorphism on $\mathbb R$ uniquely
determined by $0\leq\tilde\eta(0)<1$, $\tilde\eta(x+1) = \tilde\eta(x)+1$, and
the condition that its projects onto $\eta$.
}
\end{definition}

\begin{theorem}[Beurling-Ahlfors extension Theorem]
Let $\eta$ be an orientation preserving homeomorphism of $S^1$. Then $\eta$ is
quasisymmetric if and only if it extends to a quasiconformal homeomorphism of
the open unit disc 
$\mathbb D$ into itself.
\end{theorem}

\paragraph{$T(1)$ as a Banach manifold.}

Recall that $\mathbb D$ denotes 
the open unit disc in $\mathbb C$. Denote by $L^{\infty}(\mathbb
D)$ the complex Banach space of bounded Beltrami differentials on $\mathbb
D$.  One way to construct the universal Teichm\"uller space is the
following. Denote by $L^{\infty}(\mathbb
D)_1$ the unit ball in $L^{\infty}(\mathbb
D)$. By Ahlfors-Bers theorem, for any $\mu\in L^{\infty}(\mathbb
D)_1$, one can 
consider the unique quasiconformal mapping 
$w_{\mu}~: \mathbb D \rightarrow
\mathbb D$ which fixes $-1, -i$ and $1$ and satisfies the Beltrami equation on
$\mathbb D$
$$
\frac{\partial}{\partial\overline{z}}\omega_\mu
=\mu\frac{\partial}{\partial z}\omega_\mu.
$$
 Therefore one can define the following
equivalence relation on $L^{\infty}(\mathbb
D)_1$. For $\mu$, $\nu\in L^{\infty}(\mathbb
D)_1$, set $\mu \sim \nu$ if $w_\mu{|S^1} = w_\nu|S^1$. The universal
Teichm\"uller space is defined by the quotient space
$$
T(1) = L^{\infty}(\mathbb
D)_1/\sim.
$$
\begin{theorem}[(\cite{Le1987})]
The space $T(1)$ has a unique structure of complex Banach manifold such that
the projection map $\Phi~: L^{\infty}(\mathbb
D)_1 \rightarrow T(1)$ is a holomorphic submersion.
\end{theorem}
The differential of $\Phi$ at the origin $D_0\Phi~: L^\infty(\mathbb
D)\rightarrow T_{[0]}T(1)$ is a complex linear surjection and induces a
splitting of $L^\infty(\mathbb
D)$ into (\cite{TaTe2004})~:
$$
 L^\infty(\mathbb
D) = \operatorname{Ker}D_0\Phi \oplus \Omega_{\infty}(\mathbb D),
$$
where $\Omega^{\infty}(\mathbb D)$ is the Banach space of bounded harmonic
Beltrami differentials on $\mathbb D$ defined by
$$
\Omega_{\infty}(\mathbb D) := \left\{\mu\in L^\infty(\mathbb D)~|~\mu(z)=
(1-|z|^2)^2 \overline{\phi(z)},\quad 
\phi~~\text{holomorphic on}~~\mathbb D~\right\}.
$$

\paragraph{$T(1)$ as a group.}

By the Beurling-Ahlfors extension theorem, a quasiconformal mapping on
$\mathbb D$ extends to a quasisymmetric homeomorphism on the unit circle.  
Therefore the following map is a well-defined bijection
$$
\begin{array}{ccc}
T(1) & \rightarrow & \operatorname{QS}(S^1)/PSU(1,1)\\
\left[\mu\right] & \mapsto & \left[w_\mu|S^1\right].
\end{array}
$$
The coset $\operatorname{QS}(S^1)/PSU(1,1)$ herits
from its identification with $T(1)$ a Banach manifold structure.
Moreover the coset $\operatorname{QS}(S^1)/PSU(1,1)$ can be identified with
the subgroup of quasisymmetric homeomorphisms fixing $-1, i$ and $1$. 
This identification allows to endow the universal Teichm\"uller space with a
group structure. Relative to this differential structure, the right
translations in $T(1)$ are biholomorphic mappings, whereas the left
translations are not even continuous in general. Consequently $T(1)$ is not a
topological group.


\paragraph{The WP-metric and the Hilbert manifold structure on $T(1)$.}
The Banach manifold $T(1)$ carries a Weil-Petersson
metric, which is defined only on a distribution of the tangent bundle
(\cite{NaVe1990}). In order to resolve this problem the idea in
\cite{TaTe2004} is to change the differentiable structure of $T(1)$.
\begin{theorem}[\cite{TaTe2004}]
The universal Teichm\"uller space $T(1)$ admits a structure of Hilbert
manifold on which the Weil-Petersson metric is a right-invariant strong
hermitian metric.
\end{theorem}
For this Hilbert manifold structure, the 
tangent space at $[0]$ in $T(1)$ is isomorphic to
$$
\Omega_2(\mathbb D) := \left\{\mu(z)=
(1-|z|^2)^2 \overline{\phi(z)}, \quad\phi~~\text{holomorphic on}~~\mathbb
 D,\quad \|\mu\|_2< \infty~\right\},
$$
where $\|\mu\|_2^2 = \int\!\int_{\mathbb D}|\mu|^2\rho(z)d^2z$
is the $L^2$-norm of $\mu$ with respect to the hyperbolic metric of the
Poincar\'e disc $\rho(z)d^2z = 4(1-|z|^2)^{-2}d^2z$.
The Weil-Petersson metric on $T(1)$ is given at the tangent space 
at $[0]\in T(1)$ by 
$$
\langle\mu, \nu \rangle_{WP} := 
\int\!\!\!\int_{\mathbb D}~\mu\,\overline{\nu}\,\rho(z)d^2z
$$
With respect to this Hilbert manifold structure, $T(1)$ admits uncountably
many connected components. For this Hilbert manifold structure, the identity
component $T_0(1)$ of $T(1)$ is a topological group. Moreover 
the pull-back of the Weil-Petersson metric on the 
quotient space $\operatorname{Diff}_+(S^1)/\operatorname{PSU}(1,1)$ is given at $[\text{Id}]$ by
$$
h_{WP}([\text{Id})([u],[v])=2\pi\sum_{n=2}^\infty n(n^2-1)u_n\overline{v_n}.
$$
Hence  $T_0(1)$ of $T(1)$ 
can be seen as the completion of
$\operatorname{Diff}_+(S^1)/\operatorname{PSU}(1,1)$ for the $H^{3/2}$-norm.
This metric make $T(1)$ into a strong K\"ahler-Einstein Hilbert manifold, 
with respect to the complex structure given at $[\text{Id}]$ by the Hilbert
transform. 
The tangent space at $[\text{Id}]$ consists of Sobolev class $H^{3/2}$ vector fields modulo $\mathfrak{psu}(1,1)$. The associated Riemannian metric is given by
$$
\textrm g_{WP}([\text{Id}])([u],[v])=\pi\sum_{n\neq -1,0,1}|n|(n^2-1)u_n\overline{v_n},
$$
and the imaginary part of the Hermitian metric is the two-form
$$
\omega_{WP}([\text{Id}])([u],[v])=-i\pi\sum_{n\neq
-1,0,1}n(n^2-1)u_n\overline{v_n}.
$$
Note that $\omega_{WP}$ coincides with the Kirillov-Kostant-Souriau 
symplectic form obtained on $\operatorname{Diff}_+(S^1)/\operatorname{PSU}(1,1)$ when considered as a coadjoint orbit of the Bott-Virasoro group.

\section{The restricted Siegel disc}

\paragraph{The Siegel disc.}

Let 
$\mathcal{V} = H^{\frac{1}{2}}(S^1, \mathbb R)/\mathbb R$ be the Hilbert space 
of real valued $H^{\frac{1}{2}}$ functions with mean-value zero. The Hilbert
inner product on $\mathcal V$ is given by 
$$
\langle u, v\rangle_{\mathcal V} = \sum_{n\in\mathbb Z} |n| u_n \overline{v_n}.
$$
Endow the real Hilbert space $\mathcal V$ with the following 
complex structure  (called the Hilbert transform)
\[
\operatorname{J}\left(\sum_{n\neq 0}u_ne^{inx}\right)=i\sum_{n\neq
  0}\operatorname{sgn}(n)u_ne^{inx}.
\]
Now $\langle \cdot, \cdot \rangle_{\mathcal V}$ and $J$ are compatible in the
sense that $J$ is orthogonal with respect to $\langle \cdot, \cdot
\rangle_{\mathcal V}$. The associated symplectic form is
defined by
\[
\Omega(u,v)=\langle u , J(v)\rangle_{\mathcal V} = \frac{1}{2\pi}\int_{S^1}u(x)\partial_xv(x)dx
=-i\sum_{n\in\mathbb{Z}}nu_n\overline{v_n}.
\]
Let us consider the \textbf{complexified Hilbert space}
$\mathcal{H}:=H^{1/2}(S^1,\mathbb{C})/\mathbb{C}$ and the complex linear
extensions of $J$ and $\Omega$ still denoted by the same letters.
Each element $u\in \mathcal{H}$ can be written as
\[
u(x)=\sum_{n\in\mathbb{Z}}u_ne^{inx}\quad\text{with}
\quad u_0=0\quad\text{and}\quad \sum_{n\in\mathbb{Z}}|n||u_n|^2<\infty.
\]
The eigenspaces $\mathcal{H}_+$ and $\mathcal{H}_-$ of the operator 
$\operatorname{J}$ are the subspaces of only 
\begin{align*}
\mathcal{H}_+&=\left\{u\in\mathcal{H}\left|u(x)=\sum_{n=1}^\infty u_ne^{inx}\right.\right\}\quad\text{and}\quad
\mathcal{H}_-&=\left\{u\in\mathcal{H}\left|u(x)=\sum_{n=-\infty}^{-1} u_ne^{inx}\right.\right\},
\end{align*}
and one has the Hilbert decomposition 
 $\mathcal{H} = \mathcal{H}_+\oplus\mathcal{H}_{-}$
into the sum of closed orthogonal subspaces.
\textbf{The Siegel disc} associated with $\mathcal H$ is defined by
$$
\mathfrak{D}(\mathcal H) := \{Z\in L(\mathcal{H}_-,\mathcal{H}_+)\mid \Omega(Zu,v)=\Omega(Zv,u),\,\forall\,u,v\in\mathcal{H}_-\quad\text{and}\quad I-Z\bar Z>0\},
$$
where, for $A\in L(\mathcal{H}_+,\mathcal{H}_+)$, the notation
$A>0$ means $\langle A(u),u \rangle_{\mathcal{H}}>0$, for all
$u\in\mathcal{H}_+, u\neq 0$. For $A\in L(\mathcal{H}_+,\mathcal{H}_+)$, 
define 
$$
\overline{A}(u) := \overline{A(\bar{u})}, \quad\text\quad A^T := (\bar{A})^*.
$$
It follows easily that $\mathfrak{D}(\mathcal H)$ can be written as
$$
\mathfrak{D}(\mathcal H) := 
\{Z\in L(\mathcal{H}_-,\mathcal{H}_+)\mid Z^T = Z,\,
\forall\,u,v\in\mathcal{H}_-\quad\text{and}\quad I-Z\bar Z>0\}.
$$
The restricted Siegel disc associated with $\mathcal H$ is by definition
$$
\mathfrak{D}_{\rm{res}}(\mathcal{H})
:=\{Z\in \mathfrak{D}(\mathcal{H})\mid Z\in L^2(\mathcal{H}_-,\mathcal{H}_+)\}.
$$

\paragraph{The restricted Siegel disc as an homogeneous space.}

Consider the symplectic group $\operatorname{Sp}(\mathcal V, \Omega)$ of
bounded linear maps on $\mathcal V$ which preserve the symplectic form
$\Omega$
$$
\operatorname{Sp}(\mathcal{V},\Omega)=\{a\in\operatorname{GL}(\mathcal{V})\mid \Omega(au,av)=\Omega(u,v),\;\;\text{for all $u,v\in\mathcal{V}$}\}.
$$
The restricted symplectic group $\operatorname{Sp}_{\rm res}(\mathcal{V},\Omega)$ is by definition the intersection of the
symplectic group with the restricted general linear group defined by
$$
\operatorname{GL}_{\rm{res}}(\mathcal{H},\mathcal{H}_+)=\left\{g\in\operatorname{GL}(\mathcal{H})\mid
  \left[d,g\right]\in L^2(\mathcal{H})\right\},
$$ 
 where $d := i(p_+ - p_-)$ and $p_\pm$ is the orthogonal projection onto
 $\mathcal H_\pm$. Using the block decomposition with respect to the
 decomposition $\mathcal{H} = \mathcal{H}_+\oplus\mathcal{H}_{-}$, one gets
$$
\operatorname{Sp}_{\rm{res}}(\mathcal{V},\Omega):=\left\{\left.\left(
\begin{array}{cc}
g            &h\\
\bar{h} &\bar{g}\end{array}
\right)\in\operatorname{GL}(\mathcal{H})\right| 
h\in L^2(\mathcal{H}_-,\mathcal{H}_+), gg^* - hh^* = I, gh^T = hg^T\right\}.
$$
\begin{proposition}
The restricted symplectic group acts transitively on the restricted Siegel
disc by
$$
\operatorname{Sp}_{\rm res}(\mathcal{V},\Omega)
\times\mathfrak{D}_{\rm res}(\mathcal{H})
\longrightarrow\mathfrak{D}_{\rm res}(\mathcal{H}),\quad \left(
\left(
\begin{array}{cc}
g            &h\\
\bar{h} &\bar{g}\end{array}
\right),Z\right)\longmapsto (gZ+h)(\bar{h}Z+\bar{g})^{-1}.
$$
The isotropy group of $0\in \mathfrak{D}_{\rm res}(\mathcal{H})$ is the
unitary group $\operatorname U(\mathcal H_+)$ of $\mathcal H_+$, and the
restricted Siegel disc is diffeomorphic as Hilbert manifold to the homogeneous
space $\operatorname{Sp}_{\rm res}(\mathcal{V},\Omega)/ U(\mathcal H_+)$.
\end{proposition}
On the space $\{A \in L^2(H_-, H_+)~\mid~A^T = A\}$ consider the following
Hermitian inner product
$$
\operatorname{Tr}(V^*U) = \operatorname{Tr}(\bar{V}U). 
$$
Since it is invariant under the isotropy group of $0\in\mathfrak{D}_{\rm
  res}(\mathcal{H})$, it extends to an $\operatorname{Sp}_{\rm
  res}(\mathcal{V},\Omega)$-invariant Hermitian metric $h_{\mathfrak D}$.
\begin{remark}{\rm
In the construction above, replace $\mathcal V$ by $\mathbb R^2$ endowed with
its natural symplectic structure.  The corresponding Siegel disc is nothing
but the open unit disc $\mathbb{D}$. The action of
$\operatorname{Sp}(2,\mathbb{R})=\operatorname{SL}(2,\mathbb{R})$ is the
standard action of $\operatorname{SU}(1,1)$ on $\mathbb{D}$ given by
$$
z\in\mathbb{D}\longmapsto\frac{az+b}{\bar{b}z+\bar{a}}\in\mathbb{D},\quad |a|^2-|b|^2=1,
$$
and the Hermitian metric obtained on $\mathbb D$ 
is given by the hyperbolic metric
$$
h_\mathfrak{D}(z)(u,v)=\frac{1}{(1-|z|^2)^2}u\bar{v}.
$$
Therefore,
$\mathfrak{D}_{\rm res}(\mathcal{H})$ can be seen as 
an infinite-dimensional generalization
of the Poincar\'e disc.}
\end{remark}

\section{The period mapping}

The following theorems answer
the second question adressed in the first section.
\begin{theorem}[Theorem~3.1 in \cite{NaSu1995}]
For $\phi$ a orientation preserving homeomorphism and any $f\in\mathcal V$, 
set
by $V_{\phi}f = f \circ \varphi - \frac{1}{2\pi}\int_{S^1} f\circ \varphi$.
Then $V_{\phi}$ maps $\mathcal V$ into itself iff $\phi$ is quasisymmetric.
\end{theorem}

\begin{theorem}[Proposition~4.1 in \cite{NaSu1995}]
The group 
$\operatorname{QS}(S^1)$ of quasisymmetric
homeomorphisms of the circle acts on the right by symplectomorphisms 
on $\mathcal H =H^{1/2}(S^1,\mathbb{C})/\mathbb{C}$ by
\[
V_{\phi}f = f \circ \varphi - \frac{1}{2\pi}\int_{S^1} f\circ \varphi, 
\]
$\varphi \in \operatorname{QS}(S^1)$, $f\in\mathcal H$.
\end{theorem} 
Consequently this action defines a map $\Pi:
\operatorname{QS}(S^1)\rightarrow\operatorname{Sp}(\mathcal V, \Omega)$. Note
that the operator $\Pi(\varphi)$ preserves the subspaces $\mathcal H_+$ and
$\mathcal H_-$ iff $\varphi$ belongs to $\operatorname{PSU}(1,1)$. The
resulting map (Theorem~7.1 in \cite{NaSu1995})
is an injective equivariant holomorphic immersion
$$
\Pi~: T(1) = \operatorname{QS}(S^1)/\operatorname{PSU}(1,1) \rightarrow \operatorname{Sp}(\mathcal V, \Omega)/\operatorname{U}(H_+)\simeq\mathfrak{D}(\mathcal H)
$$ 
called the \textbf{period mapping} of $T(1)$. The Hilbert version of the
period mapping is given by the following
\begin{theorem}[\cite{TaTe2004}]
For $[\mu]\in T(1)$, $\Pi([\mu])$ belongs to the restricted Siegel disc if and
only if $[\mu]\in T_0(1)$. Moreover the pull-back of the natural K\"ahler
metric on $\mathfrak{D}_{\rm res}(\mathcal{H})$ coincides, up to a constant
factor, with the Weil-Petersson metric on $T_0(1)$.
\end{theorem}


\bibliographystyle{new}

\end{document}